\documentclass{elsart}

\usepackage{natbib}

\usepackage{graphicx}

\usepackage{amssymb}

\usepackage{rotating}

\begin{document}

\begin{frontmatter}



\title{Latest results on Jovian disk X-rays from {\it XMM-Newton}}


\author[MSSL]{G. Branduardi-Raymont},
\author[India]{A. Bhardwaj},
\author[MSFC]{R. F. Elsner},
\author[SWRI]{G. R. Gladstone},
\author[MSSL]{G. Ramsay},
\author[VILSPA]{P. Rodriguez},
\author[MSSL]{R. Soria},
\author[ANN]{J. H. Waite, Jr},
\author[KENT]{T. E. Cravens}

\address[MSSL]{Mullard Space Science Laboratory, University College London,
              Holmbury St Mary, Dorking, Surrey RH5 6NT, UK}
\address[India]{Space Physics Laboratory, Vikram Sarabhai Space Centre, 
                Trivandrum 695022, India}
\address[MSFC]{NASA Marshall Space Flight Center, NSSTC/XD12,
              320 Sparkman Drive, Huntsville, AL 35805, USA}
\address[SWRI]{Southwest Research Institute, P. O. Drawer 28510,
              San Antonio, TX 78228, USA}
\address[VILSPA]{XMM-Newton SOC, Apartado 50727, Villafranca, 28080 Madrid,
              Spain}
\address[ANN]{University of Michigan, Space Research Building,
              2455 Hayward, Ann Arbor, MI 48109, USA}
\address[KENT]{Department of Physics and Astronomy, University of Kansas, 
              Lawrence, KS 66045, USA}

\begin{abstract}

We present the results of a spectral study of the soft X-ray emission
(0.2$-$2.5 keV) from low-latitude (`disk') regions of Jupiter. The data were 
obtained during two observing campaigns with {\it XMM-Newton} in April and 
November 2003. While the level of the emission remained approximately the same 
between April and the first half of the November observation, the second part
of the latter shows an enhancement by about 40\% in the 0.2$-$2.5 keV flux.
A very similar, and apparently correlated increase, in time and scale, was
observed in the solar X-ray and EUV flux.

The months of October and November 2003 saw a period of particularly intense
solar activity, which appears reflected in the behaviour of the soft X-rays
from Jupiter's disk. The X-ray spectra, from the {\it XMM-Newton} EPIC CCD 
cameras, are all well fitted by a coronal
model with temperatures in the range 0.4$-$0.5 keV, with additional 
line emission from Mg XI (1.35 keV) and Si XIII (1.86 keV): these are 
characteristic lines of solar X-ray spectra at maximum activity and 
during flares.

The {\it XMM-Newton} observations lend further support to the theory that 
Jupiter's disk X-ray emission is controlled by the Sun, and may be produced in 
large part by scattering, elastic and fluorescent, of solar X-rays in the 
upper atmosphere of the planet.

\end{abstract}


\end{frontmatter}

\section{Introduction}
\label{}

The current generation of X-ray observatories, with their greatly improved
spatial resolution ({\it Chandra}) and sensitivity ({\it XMM-Newton}),
coupled to moderate (CCD) to high (gratings) spectral resolution,
have made it feasible for the first time to study solar system objects
in detail. X-ray observations of planets, satellites and comets, coupled with
solar X-ray studies and solar wind and magnetospheric measurements {\it in 
situ}, are being used to extend our understanding of the photon and particle 
processes taking place all over the solar system, of the interactions 
between the Sun and solar system bodies, and ultimately of the effects 
solar activity may have on our own Earth (for a review see Bhardwaj et al.,
2006b, this issue). 

Jupiter has a particularly complex magnetospheric environment, which is 
governed by the fast rotation of the giant planet, and by the presence 
of Io and its dense plasma torus. This made it an interesting target since
the earliest attempts of solar system X-ray studies: Jupiter was first 
detected at X-ray energies with the {\it Einstein} observatory \citep{Metz}; 
later studies with {\it ROSAT} (Waite et al., 1994, 1997) established 
the presence of two distinct
types of X-ray emission from the planet: `auroral', from regions near the 
magnetic poles, and `disk' emission, from lower latitudes. The present
paper focuses on the disk emission; recent detailed studies 
of the Jovian X-ray aurorae can be found in Branduardi-Raymont et al. (2004,
2006a,b) and \citet{els05}, and references therein.

The first attempt to explain the origin of Jupiter's disk emission 
\citep{Waite97} was based on energetic ion precipitation, the same mechanism 
invoked to account for the auroral emissions also seen by {\it ROSAT}. 
However, a general decrease in the overall X-ray brightness of Jupiter over 
the years 1994-1996 was found to be coincident with a similar decay
in solar activity, as characterised by the solar 10.7 cm flux; the 
X-ray flux from the Jovian disk also revealed an interesting dependence 
on local time, with the X-ray bright limb coinciding with the bright visible
limb \citep{gla98}. These two facts combined suggested that the planet's disk 
emission may be controlled by the Sun to some extent. Indeed \citet{mau00}
were able to demonstrate that scattering of solar X-rays in Jupiter's
atmosphere may well play a role in generating its equatorial emission.
They modelled this component of solar origin with a combination of 
elastic scattering by atmospheric neutrals, and, to a smaller degree, 
fluorescent scattering of C K-shell X-rays on methane molecules. 

{\it Chandra} High Resolution Camera (HRC) observations of Jupiter in Dec. 
2000 \citep{gla02} not only revealed surprising characteristics in its auroral 
emission, but gave us the sharpest X-ray view of the planet yet, clearly 
separating the bright auroral and the essentially uniform disk emissions (see 
Figure 28 of Bhardwaj et al., 2006b, this issue). However, no information on
the spectral characteristics of the emissions could be obtained from the 
HRC data. Jupiter was observed again in Feb. 2003 with the {\it Chandra} 
Advanced CCD Imaging Spectrometer (ACIS), and the HRC, and this time spectra 
were acquired of both auroral \citep{els05} and disk \citep{bhar06a} emissions.
Recently \citet{cra06} have produced calculations of the scattering albedo 
for soft X-rays from the outer planets and have compared them against the 
ACIS spectra of Jupiter's disk. The 
conclusion is that indeed the soft X-ray emissions of Jupiter (and Saturn)
can largely be explained by scattering and fluorescence of solar X-rays.

A study of Jupiter's soft X-ray emission from the first {\it XMM-Newton} 
observation of the planet in Apr. 2003 \citep{bra04} clearly
demonstrated the different characteristics of the auroral and disk spectra: 
while charge exchange was found most likely to be responsible for the auroral
soft X-ray production, a coronal (i.e. optically thin collisional plasma) 
model best fitted the disk emission, 
giving support to the solar X-ray scattering hypothesis. In the following 
we describe the results of a second {\it XMM-Newton} observation, carried out 
in Nov. 2003 (a preliminary account of this can be found in 
Branduardi-Raymont et al., 2006a), and we compare them with those from 
a re-analysis of the Apr. 2003 data. 


\section{{\it XMM-Newton} observations}
\label{}

{\it XMM-Newton} \citep{jan01} observed Jupiter for two consecutive 
spacecraft revolutions (0726 and 0727; a total of 245 ks) between Nov. 25, 
23:00 and Nov. 29, 12:00, i.e. for more than twice the duration of the Apr.
2003 observation (110 ks, Branduardi-Raymont et al., 2004). 
As on that occasion the EPIC-MOS \citep{turner} 
and -pn \citep{struder} cameras (with a field of view of 30' diameter)
were operated in Full Frame and Large Window mode respectively,
and the RGS instrument \citep{denHerder} in Spectroscopy.
The filter wheel of the OM telescope \citep{Mason} was kept in
the blocked position because the optical brightness of Jupiter is above the
safe limit for the instrument, so no OM data were collected; also to minimise
the risk of optical contamination in the X-ray data the EPIC cameras were 
used with the thick filter. Jupiter's motion on the sky (11 and 16''/hr
in Apr. and Nov. 2003 respectively) required several pointing trims during the 
long observations so that the target would not move out of the 
cameras' central CCD chip, and to avoid worsening the RGS spectral resolution.
During both observations the planet's path on the sky was roughly along the
RGS dispersion direction, so that good separation of the two auroral 
spectra could be achieved. The planet's disk diameter was 38'' in Apr. and 36''
in Nov. 2003. 

The data were processed and analysed with the {\it XMM-Newton} Science
Analysis Software (SAS) (see SAS User's Guide at
http://xmm.vilspa.esa.es/extern- al/xmm\_user\_support/documentation). Photons
collected along Jupiter's path during the pointings were referred to the
centre of the planet's disk so that images, lightcurves and average spectra
could be constructed in the planet's reference system. Exclusion of data
affected by high particle background at the end of both spacecraft
revolutions in Nov. 2003 leaves a total of 210 ks of good quality data
for analysis (this total was 80 ks in Apr. 2003). Figure 1 displays the
0.2$-$2.0 keV image of Jupiter obtained combining the Nov. 2003 data for all 
{\it XMM-Newton} EPIC CCD cameras. 

\section{Temporal analysis}
\label{}

Unlike in Apr. 2003 \citep{bra04}, the EPIC lightcurve of X-ray events 
extracted from the equatorial regions of Jupiter in Nov. 2003 presents evidence
for variability, with a smooth visible increase in flux from beginning to end 
by $\sim$40\%. A similar increase is present in the solar EUV and X-ray fluxes 
over the same period (see Figure 2, taken from Bhardwaj et al., 2005). 
Moreover, a large solar X-ray flare 
taking place on the Jupiter-facing side of the Sun (at 2.4 days into the 
observation; see Figure 2, this paper, and Figure 30 of Bhardwaj et al., 
2006b, this issue) appears
to have a corresponding feature in the Jovian X-rays. Both these facts
support the hypothesis that Jupiter's disk emission is predominantly 
scattered solar X-rays from the planet's upper atmosphere and that it is
directly controlled by solar irradiation. It is interesting to note
that the months of Oct. and Nov. 2003 corresponded to a period of 
particularly intense solar activity, of the kind previously observed
during the decay phases of a solar cycle, and much stronger than that 
in Apr. 2003.

\section{Spectral study}
\label{}

The spectral differences between the disk and the auroral X-ray emissions
of Jupiter, already revealed by the Apr. 2003 {\it XMM-Newton} observation
(Brandu-ardi-Raymont et al., 2004), are very evident when we construct spectral 
maps of the planet in narrow energy bands using the Nov. 2003 EPIC data 
(Figure 3); the bright, well separated aurorae shine at the lowest energies 
(spectral bands centred on the O VII, 0.57 keV, and O VIII, 0.65 keV, emission 
lines), while the more uniform and round-shaped emission from the disk emerges 
clearly at higher energies (around the Fe XVII and Mg XI ionic transitions, 
at 0.7$-$0.8 and $\sim$1.4 keV respectively). 

When extracting Jupiter's auroral and disk spectra from the EPIC data the
blurring effect introduced by the {\it XMM-Newton} Point Spread Function (PSF,
$\sim$15'' Half Energy Width) has to be taken into account, and some form
of `de-mixing' of the X-ray events from the different parts of the planet
needs to be carried out. This was done by subtracting appropriate fractions of 
disk and auroral emissions from the aurorae and the disk spectra respectively. 
These fractions were established by convolving the {\it XMM-Newton} PSF with 
the summed {\it Chandra} ACIS and HRC surface brightness distributions of 
Jupiter observed in Feb. 2003 (displayed in Figure 4; Elsner et al., 2005) 
for the regions used in the 
extraction. These regions are shown in Figures 1 and 4; the rectangular box at 
the centre, from which the disk spectrum was extracted, has sides of 51.7'' 
and 14.9'', and covers 32\% of the circle enclosing the {\it XMM-Newton} image 
of the whole planet (see Branduardi-Raymont et al., 2006b, for details of the
spectral extraction). 

Figure 5 shows a comparison of the EPIC disk spectra
over the band 0.2$-$10 keV from the Apr. and Nov. 2003 {\it XMM-Newton}
observations. The latter was broken up into two observing intervals
defined by the two consecutive spacecraft revolutions (0726 and 0727),
so that we have a total of three datasets of similar duration (80$-$100 ks).
The main reasons for separating the two revolutions were the variability 
in the overall level of the disk emission, which increased from beginning
to end of the Nov. observation (see Figure 2), and also the remarkable 
change observed in the auroral spectra between the two intervals 
(Branduardi-Raymont et al., 2006a,b); the latter suggests that a 
significant change took place in 
the planet's magnetospheric environment over the two halves of the observation.
The Apr. 2003 spectrum was re-extracted using the same technique as for
the Nov. observation, so that the three spectra in Figure 5 are directly 
comparable.

The horizontal green line in Figure 5 represents the predicted
level of particle background (the cosmic ray diffuse background is occulted 
by the planet's body). This, together with the systematic flattening of the 
spectra at high energies, suggests that the disk emission dominates and is
well above the particle background only up to $\sim$2.5 keV. Thus spectral
fitting was carried out in the band 0.2$-$2.5 keV. The spectra were binned at
a minimum of 20 counts per bin, so that the $\chi^2$ minimization technique
would be applicable.

Given the likely connection between the observed disk X-ray emission of 
Jupiter and the solar one, we started by considering that a solar spectrum
may be a good model to apply to the {\it XMM-Newton} data of the planet.
\citet{per00} synthesized the solar X-ray spectrum from {\it Yohkoh}/SXT
data and folded it through the instrumental responses of non-solar X-ray
observatories, to generate simulated solar spectra at the {\it ASCA} 
spectral resolution. They then derived spectral
parameters from a best fit analysis of the data, and did this for three cases: 
solar minimum, solar maximum and flaring Sun. We have taken their parameters 
as the initial ones for our fits to Jupiter's disk spectra. 
Briefly, \citet{per00} spectra were synthesized using coronal emission 
components ({\tt mekal} model in XSPEC, the standard spectral fitting code 
in X-ray astronomy): one {\tt mekal} component (kT = 0.15 keV) was sufficient
to describe the solar minimum spectrum, two (kT = 0.18 and 0.49 keV) 
were required at solar maximum and three (kT = 0.21, 1.07, 1.59 keV) 
for the flaring Sun. The appearence of the synthetic spectra changes 
significantly from the first to the third case, and this is clearly reflected 
in the simulated {\it ASCA} data (which are at a similar resolution to 
that of {\it XMM-Newton} EPIC): the spectrum becomes increasingly harder, with 
the flux rising in the 1$-$5 keV band, and emission features appear prominently
at and above 1 keV; in particular, line emission between 1 and 2 keV, 
corresponding to Mg XI and Si XIII transitions \citep{phillips}, is 
strong at solar maximum (Figure 7 of Peres et al., 2000), while the 
spectrum of the flaring Sun (their Figure 8) has a peak around 1 keV, due
to a combination of Fe XVII, Fe XXI and Ne X lines \citep{phillips}.

We first combined the spectra from the three {\it XMM-Newton} datasets in order
to establish the significance of the spectral features and characterize them
at the highest possible signal-to-noise level. The best fit is obtained with
a single {\tt mekal} model with a temperature of 0.42$^{\rm +0.01}_{\rm -0.02}$
keV and solar abundances, and the inclusion of line emission (described by a
gaussian function in XSPEC) from the Mg XI resonance transition at 1.35 keV
and from that of Si XIII at 1.86 keV. Given the moderate statistical quality
of the data, the line energies were kept fixed in the fits, in order to limit
the number of degrees of freedom. We confirm that the presence of line 
emission from both Mg and Si ions is statistically significant in the combined 
spectrum. The best fit parameters are listed in Table 1. Figure 6 displays
the data and the best fit model, together with the $\chi^2$ contributions
for all spectral bins. Indicated in the figure are the locations of the Mg XI
and Si XIII lines, as well as those of the most prominent line emission
contributions from the coronal plasma (Fe XVII at 0.7$-$0.8 keV, Fe XVII and
XXI and Ne X at $\sim$1 keV). The results of this spectral fitting reinforce
and quantify what was visually already clear from Figure 3: the
iron emission (especially the bright features around 0.7$-$0.8 keV) and that,
weaker, of Mg XI, clearly map out the X-ray brightness of the whole 
disk of Jupiter; in the spectral fitting, however, we only used X-ray events 
extracted from the low latitude regions of the planet, where a proper 
de-mixing from the auroral emission can be carried out. 

We also analysed the three {\it XMM-Newton} spectra from Apr. and Nov. 2003
separately, to check for any sign of variability beween them. The same 
combination of coronal plasma and line emission is required to represent the 
individual spectra as the combined one: the {\tt mekal} temperature lies in the
range 0.4$-$0.5 keV; the presence of Mg XI emission is clearly detected in the 
individual spectra, but that of Si XIII is less convincing. The best fit 
parameters are listed in Table 1 and the spectra and best fits are shown in 
Figure 7.
The calculations of line equivalent widths in particular suffer from large 
uncertainties: this is most likely due to the low flux observed, and thus 
the severe binning required, at the upper end of the energy range examined. 
Moreover, it must be noted that the equivalent widths listed in Table 1 are 
for the lines added to the coronal model, which already includes some Mg XI 
and Si XIII emission, as appropriate for the best fit temperature; this 
makes the level of the underlying continuum very dependent on 
the energy band over which it is estimated, and the equivalent widths (which
are essentially lower limits to the total emission from the transitions) 
particularly inaccurate.

There is evidence in all three spectra (marginally in rev. 0727) of emission,
unaccounted for, at $\sim$0.57 keV (the energy of the resonance transition of 
O VII), which may be residual contamination from the auroral regions. 
Alternative models, such as two coronal components, were tried, but produced 
statistically worse fits. 

The 0.2$-$2.5 keV fluxes listed in Table 1 have been used to calculate the 
emitted power at Jupiter after normalisation to the total area of the planet's
disk. As mentioned above, the low-latitude spectral 
extraction box (see Figure 1) is about one third in size of the circle
enclosing the planet's image, which is blurred by the {\it XMM-Newton} PSF
(15'' Half Energy Width). Assuming that the circle (of radius 27.5'') contains 
the vast majority of the X-ray emission from Jupiter, the power emitted
from the planet's full disk (excluding the aurorae) is estimated to be 
0.37, 0.38 and 0.58 GW in Apr. and Nov. (rev. 0726 and 0727) 2003 respectively.

\section{Discussion and conclusions}
\label{}

The results of the spectral analysis reported in the previous section
reinforce the idea that Jupiter's low-latitude disk X-ray emission, away from 
the auroral zones, is predominantly the consequence of the Sun's X-ray 
irradiation and of scattering in the upper atmosphere of the 
planet \citep{cra06}: the {\it XMM-Newton} spectra are very well fitted by a
coronal model with a temperature which is very similar to that observed 
at solar maximum for the Sun's X-rays. The Mg XI and Si XIII line emissions, 
also observed in Jupiter, are considered the signatures of flaring episodes on 
the Sun: the Jovian Mg XI flux was strongest in Nov. 2003, rev. 0727,
when Jupiter's disk flux was highest and flares were observed in both solar 
and Jupiter's X-ray lightcurves (see Bhardwaj et al., 2005).

However, with the available data we cannot exclude the possibility that other 
processes may play a role in this scenario: recent results \citep{bhar06a} have
confirmed the existence of a correlation between low-latitude X-ray intensity 
and surface magnetic field strength which was originally reported in 
{\it ROSAT} data \citep{gla98}, and which seemed to contradict other evidence 
supporting a solar connection. This would suggest that some phenomenon of 
magnetospheric origin is also at work at low latitudes on Jupiter.



\section{Acknowledgements}
\label{}

This work is based on observations obtained with {\it XMM-Newton}, an ESA
science mission with instruments and contributions directly funded by ESA
Member States and the USA (NASA). The MSSL authors acknowledge financial
support from PPARC.
 


 \begin{sidewaystable}
      \caption[]{Best fit parameters for the 0.2$-$2.5~keV spectra of
Jupiter's low-latitude disk region (errors are at 90\% confidence)}
         \label{fits}

\medskip 

 $
         \begin{array}{ccccccccc}
            \hline\hline
            \noalign{\smallskip}
XSPEC~component & \multicolumn{2}{c}{Apr. + Nov. 2003} &
                  \multicolumn{2}{c}{Apr.~2003}        &
                  \multicolumn{2}{c}{Nov.~2003~(0726)} &
                  \multicolumn{2}{c}{Nov.~2003~(0727)}    \\
            \noalign{\smallskip}
            \hline
            \noalign{\smallskip}
{\tt mekal} & kT^{\mathrm{a}} & Norm^{\mathrm{b}} &
              kT^{\mathrm{a}} & Norm^{\mathrm{b}} &
              kT^{\mathrm{a}} & Norm^{\mathrm{b}} &
              kT^{\mathrm{a}} & Norm^{\mathrm{b}} \\
            \noalign{\smallskip}
   & 0.42 ^{\rm +0.01}_{\rm -0.02} & 18.3^{\rm +0.7}_{\rm -0.8}
   & 0.51\pm{0.04}  & 15.8^{\rm +1.0}_{\rm -1.2}
   & 0.43\pm{0.03}  & 14.8\pm{1.3}
   & 0.39\pm{0.02}  & 23.0^{\rm +1.3}_{\rm -1.2} \\
            \noalign{\smallskip}
             \hline
            \noalign{\smallskip}
Gauss.~line^{\mathrm{c}} & Flux^{\mathrm{d}} & EW^{\mathrm{e}}
                           & Flux^{\mathrm{d}} & EW^{\mathrm{e}}
                           & Flux^{\mathrm{d}} & EW^{\mathrm{e}}
                           & Flux^{\mathrm{d}} & EW^{\mathrm{e}} \\
            \noalign{\smallskip}
1.35 & 0.67^{\rm +0.19}_{\rm -0.16} & 230^{\rm +110}_{\rm -150} &
       0.48^{\rm +0.46}_{\rm -0.36} & 320^{\rm +200}_{\rm -240} &
       0.33^{\rm +0.37}_{\rm -0.21} & 120^{\rm +370}_{\rm -120} &
       1.04^{\rm +0.34}_{\rm -0.33} & 340^{\rm +240}_{\rm -260} \\
            \noalign{\smallskip}
1.86 & 0.33^{\rm +0.11}_{\rm -0.12} & 550^{\rm +380}_{\rm -390} &
       0.41^{\rm +3.20}_{\rm -0.19} & 540^{\rm +610}_{\rm -540} &
       0.75^{\rm +2.25}_{\rm -0.20} & 1400^{\rm +9300}_{\rm -1400} &
       0.38^{\rm +0.33}_{\rm -0.17} & 620^{\rm +740}_{\rm -620} \\
            \noalign{\smallskip}
             \hline
            \noalign{\smallskip}
 \chi^2~(d. o. f.) & \multicolumn{2}{c}{147.3~(88)} &
                     \multicolumn{2}{c}{43.5~(57)}  &
                     \multicolumn{2}{c}{57.6~(56)}  &
                     \multicolumn{2}{c}{83.1~(70)} \\
           \noalign{\smallskip}
             \hline
            \noalign{\smallskip}
Flux^{\mathrm{f}}~0.2-2.5~{\rm keV} & \multicolumn{2}{c}{3.7}
                              & \multicolumn{2}{c}{3.2} 
                              & \multicolumn{2}{c}{3.0}
                              & \multicolumn{2}{c}{4.5} \\
            \noalign{\smallskip}
            \hline
         \end{array}
 $ \\

\begin{list}{}{}
\item[$^{\mathrm{a}}$] {\tt mekal} temperature in keV
\item[$^{\mathrm{b}}$] {\tt mekal} normalisation in units of
10$^{\rm -6}$ ph cm$^{\rm -2}$ s$^{\rm -1}$ keV$^{\rm -1}$
\item[$^{\mathrm{c}}$] Energy of the emission line in keV (fixed in the fits)
\item[$^{\mathrm{d}}$] Total flux in the line in units of
10$^{\rm -6}$ ph cm$^{\rm -2}$ s$^{\rm -1}$
\item[$^{\mathrm{e}}$] Line equivalent width in eV
\item[$^{\mathrm{f}}$] Flux in units of 10$^{\rm -14}$ erg cm$^{\rm -2}$ 
s$^{\rm -1}$
\end{list}
            \smallskip
   \end{sidewaystable}

\eject

\begin{figure}
\centering
\includegraphics[angle=-90,scale=0.5]{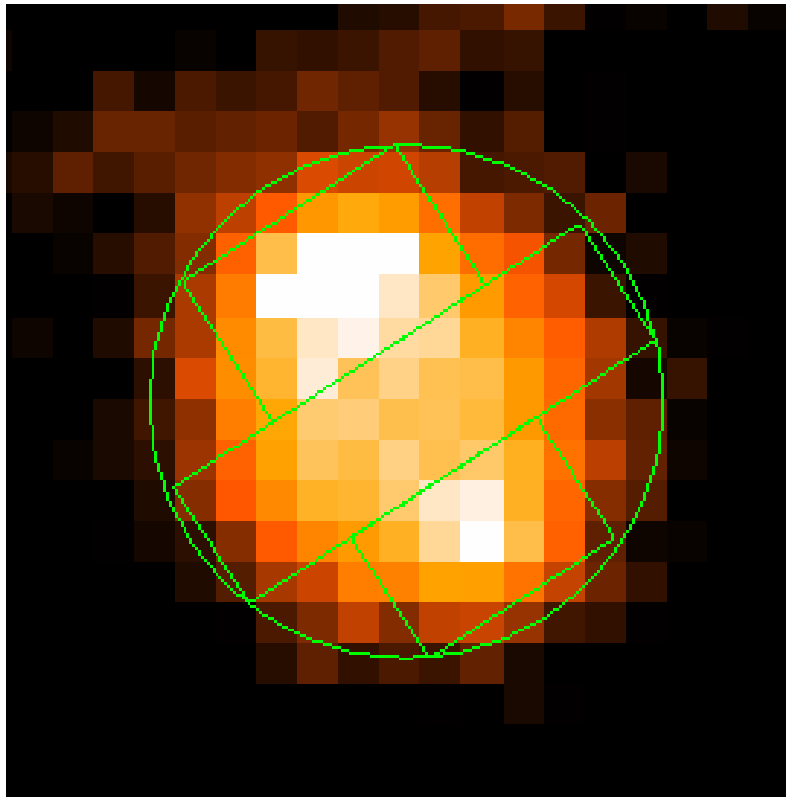}
\caption{Jupiter's image ($\sim$1.4' side, 0.2$-$2.0 keV) from the Nov. 2003 
{\it XMM-Newton} EPIC cameras data; North is to the top,
East to the left. Superposed are the regions used to extract auroral and 
low-latitude disk lightcurves and spectra.
\label{Fig1}
}
\end{figure}

\begin{figure}
\centering
\includegraphics[angle=-90,scale=0.5]{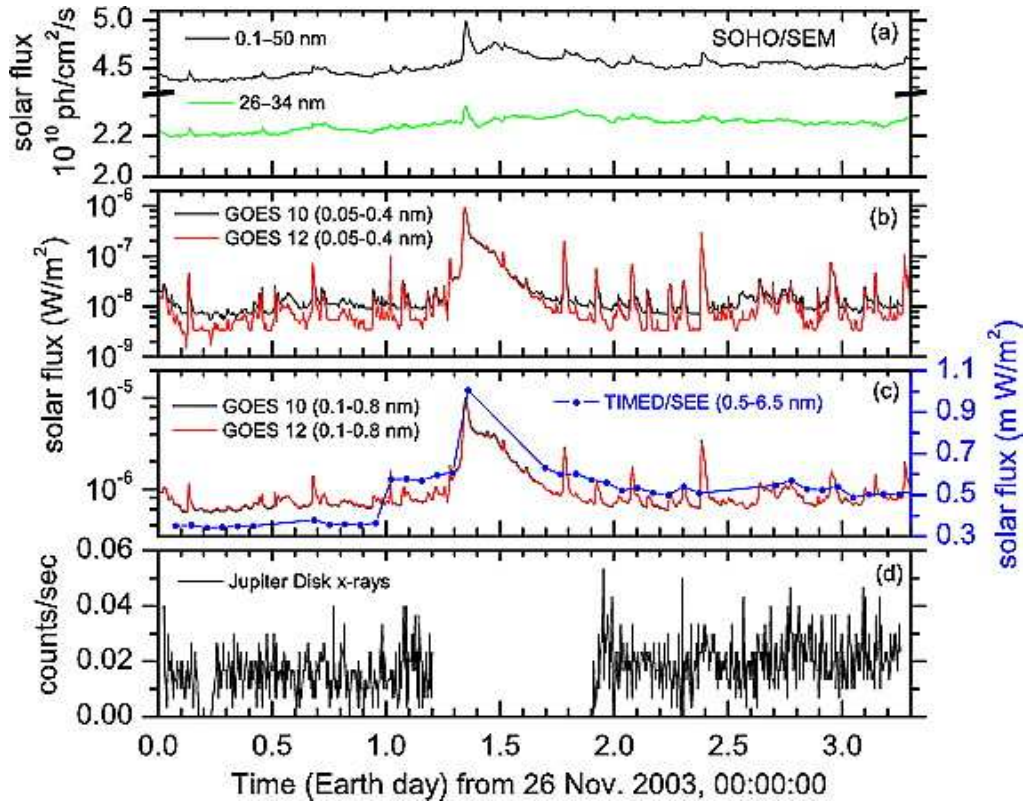}
\caption{Comparison (from Bhardwaj et al., 2005) of Jupiter's disk X-ray 
lightcurve with that of solar X-rays (scaled to 1 AU), in 5 min bins, unless 
otherwise specified, for the Nov. 2003 {\it XMM-Newton} observation.
(a): Solar SOHO/SEM 0.1$-$50 nm (0.02$-$12.4 keV) and 26$-$34 nm
(0.04$-$0.05 keV) (note the break in the y-axis).
(b): Solar 0.05$-$0.4 nm (3$-$25 keV) flux from GOES 10 and 12.
(c): Solar 0.1$-$0.8 nm (1.6$-$12.4 keV) flux from GOES 10 and 12 (practically
indistinguishable), and solar 0.5$-$6.5 nm (0.2$-$2.5 keV) flux form TIMED/SEE.
(d) {\it XMM-Newton} EPIC 0.2$-$2.0 keV lightcurve for Jupiter's 
equatorial regions, shifted by -4948 s to account for travel time delay
between Sun-Jupiter-Earth and Sun-Earth. The small gap at 0.2 days into the 
observation is due to loss of telemetry and the longer one between 1.2 and 
1.9 days to the spacecraft's perigee passage.
\label{Fig2}
}
\end{figure}

\begin{figure}
\centering
\includegraphics[angle=-90,scale=0.18]{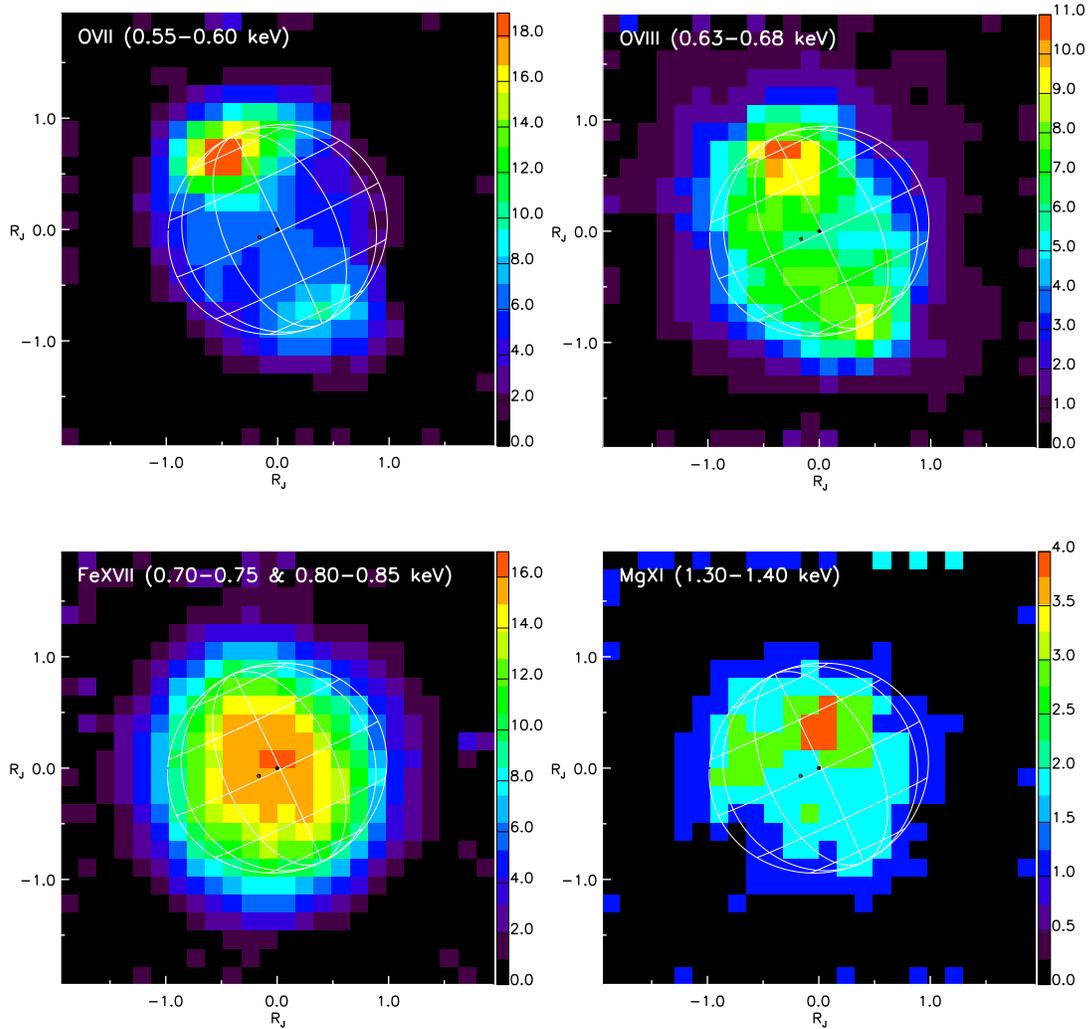}
\caption{Smoothed {\it XMM-Newton} EPIC images of Jupiter in narrow
spectral bands. The spectral ranges are centered, from top left, clockwise,
on the ionic transitions of O VII (0.55$-$0.60 keV), O VIII (0.63$-$0.68 keV),
Fe XVII (0.70$-$0.75 and 0.80$-$0.85 keV) and Mg XI (1.30$-$1.40 keV).
The colour scale bar is in units of EPIC counts. A graticule showing Jupiter's
orientation with 30$^{\rm o}$ intervals in latitude and longitude is overlaid.
The circular mark indicates the sub-solar point; the sub-Earth point is at
the centre of the graticule.
\label{Fig3}
}
\end{figure}

\begin{figure}
\centering
\includegraphics[angle=0,scale=1.2]{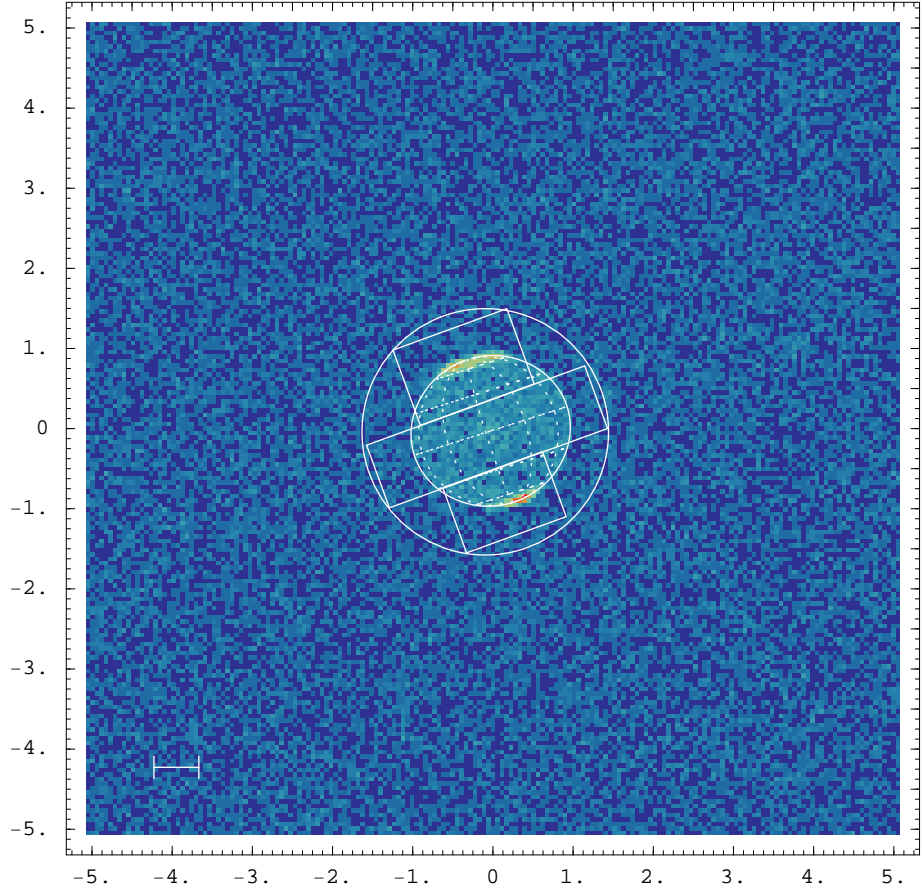}
\caption{Summed {\it Chandra} ACIS and HRC surface brightness distribution
observed for Jupiter during the Feb. 2003 observation; superposed are the
extraction regions used for the {\it XMM-Newton}~data analysis. The axes are
in units of Jupiter's equatorial radius at the time of the Nov. 2003
{\it XMM-Newton} observation. The white scale bar in the lower left has a
length of 10''. The extraction regions have the same areas as those used
for the {\it XMM-Newton}~data, but are rotated by 20$^{\rm o}$ rather than
33$^{\rm o}$ to compensate for the different orientation of the planet at the
two epochs.
\label{Fig4}
}
\end{figure}

\begin{figure}
\centering
\includegraphics[angle=-90,scale=0.5]{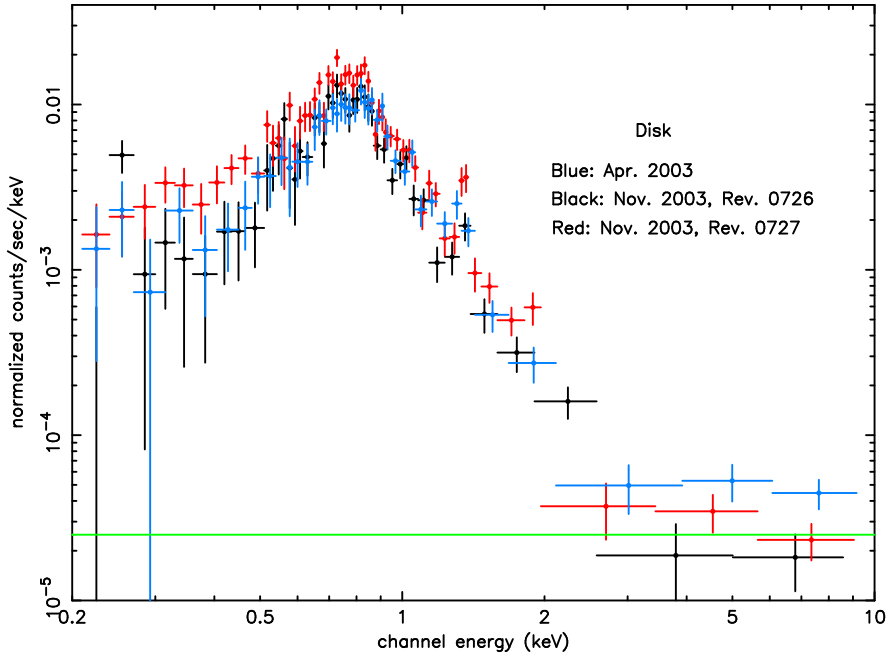}
\caption{EPIC spectra of the low-latitude disk emission for two consecutive
{\it XMM-Newton} revolutions, 0726 (black) and 0727 (red) in Nov. 2003, 
and for the Apr. 2003 observation (blue). The green horizontal line indicates
the predicted level of particle background.
\label{Fig5}
}
\end{figure}

\begin{figure}
\centering
\includegraphics[angle=-90,scale=0.4]{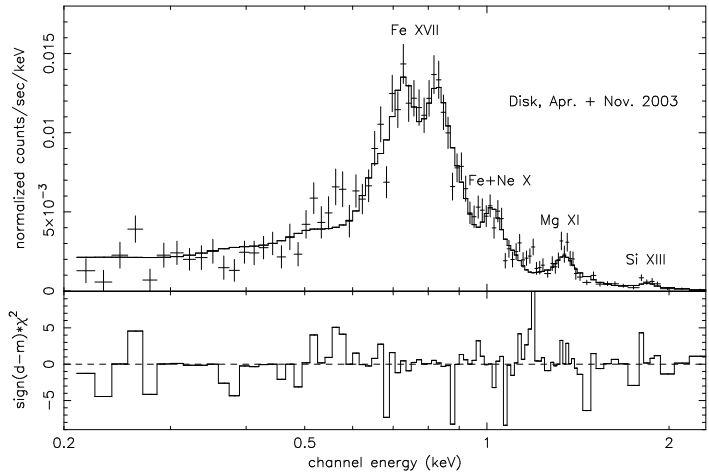}
\caption{Jupiter's low-latitude disk spectrum (crosses) and best fit model
(histogram) for the {\it XMM-Newton} EPIC observations in Apr. and Nov. 2003 
combined. At the bottom the $\chi^2$ contribution for each spectral bin is
plotted. Labels indicate the locations of the emission features from Fe XVII at
0.7$-$0.8 keV and of the blend of Fe XVII, FeXXI and Ne X at $\sim$1 keV, all 
due to the coronal plasma spectral component. Additional line emission from 
Mg XI and Si XIII (likely to be due to solar activity) is also indicated.
\label{Fig6}
}
\end{figure}

\begin{figure}
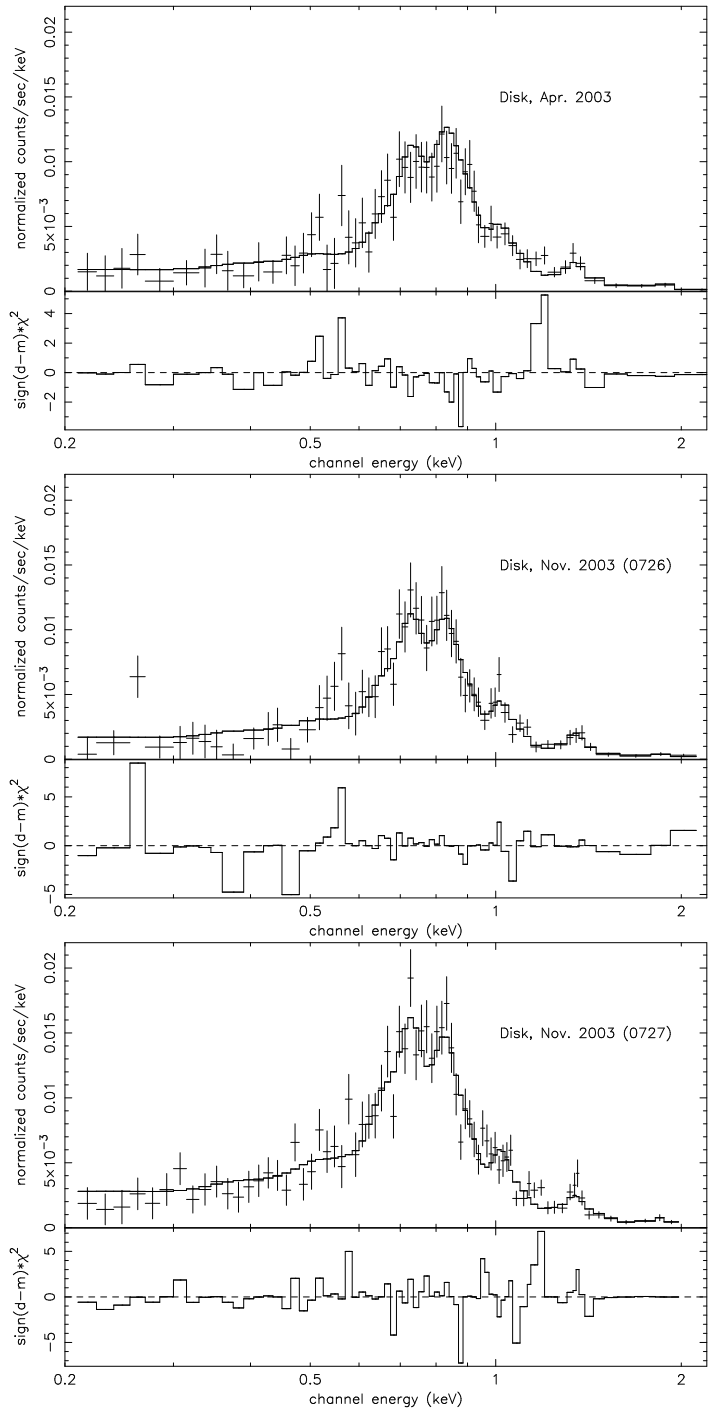

\centering
\includegraphics[angle=-90,scale=0.4]{apr03_equat_solar_max_lines_fix.ps}
\includegraphics[angle=-90,scale=0.4]{0726_equat_solar_max_lines_fix.ps}
\includegraphics[angle=-90,scale=0.4]{0727_equat_flaring_sun_fix.ps}
\caption{Best fits to Jupiter's low-latitude disk spectra from the individual 
{\it XMM-Newton} observations in Apr. (top) and Nov. 2003 (middle and bottom 
panels). Data are shown by crosses and models by histograms. At the bottom 
of each panel is plotted the $\chi^2$ contribution for each spectral bin. 
Emission from Mg XI at 1.35 keV is very clear in all the plots, while the 
weaker contribution from Si XIII at 1.86 keV is not statistically significant
in these individual datasets.
\label{Fig7}
}
\end{figure}

\end{document}